\documentclass[aps, prstab, reprint, showpacs, amsmath, amssymb]{revtex4-1}

\usepackage{amsmath,amscd}
\usepackage{bm}% bold math
\usepackage[english]{babel}
\usepackage{graphicx}% Include figure files
\usepackage{soul}
\usepackage[usenames]{color}
\begin{document}
\title{ Cherenkov Wakefield Radiation from an Open End of a Three-Layer Dielectric Capillary }

\author{Sergey N. Galyamin}
\email{s.galyamin@spbu.ru}
\affiliation{Saint Petersburg State University, 7/9 Universitetskaya nab., St. Petersburg, 199034 Russia}
%\author{Viktor V. Vorobev}
%\affiliation{Technical University of Munich, 80333 Munich, Germany}

\date{\today}

\begin{abstract}
Modern trends in beam-driven radiation sources involve interaction of Cherenkov wakefields with open-ended circular waveguide structures having complicated dielectric lining, with a three-layer dielectric capillary recently offered for reducing the radiation divergency being a representative example~\cite{Jiang2020}.
This paper presents rigorous approach allowing analytical description of electromagnetic processes occuring when the described structure is excited by single waveguide mode.
In other words, corresponding canonical waveguide diffraction problem is considered in rigorous formulation.
This is continuation of our recent paper~\cite{GV2022} where a simpler case of a two-layer dielectric filling has been considered.
Here we use the same analytical approach based on Wiener-Hopf-Fock technique and deal with more complicated case of a three-layer dielectric lining.

%
%As a specific example directly applicable to beam-driven radiation sources based on dielectric-lined capillaries, diffraction of a slow TM symmetrical mode at the open end of the described waveguide is considered.
%A series of such modes forms the wakefield (Cherenkov radiation field) generated by a charged particle bunch during its passage along the vacuum channel axis.
%Calculated S-parameters were compared with those obtained from COMSOL simulation and an excellent agreement was shown.
\end{abstract}

\maketitle

\section{Introduction\label{sec:intro}}

Dielectric-lined open-ended waveguide structures are considered nowadays as extremely promising for a variety of applications based on Cherenkov effect.
%
%Cherenkov radiation (CR) has been initially discovered with fast electrons traversing dielectric medium and emitting radiation in the visible region of electromagnetic spectrum~\cite{Ch37}. 
%Through decades, CR has been successfully used for a variety of applications in high-energy physics connected mainly with particle detection~\cite{Jb, Zrb}.
%Today considerable advances have been reached in implementation of CR effect in a variety of contemporary accelerator-based technologies.
%
In the context of the present paper, one can mention certain success in both dielectric wakefield acceleration~\cite{Nanni2015, OShea16, WangAntipov17, JingAntipov18, Pacey2020, Tang2021}%
% where CR in the form of a wakefield with up to GV per meter magnitude and Terahertz (THz) frequencies can be generated by high-quality relativistic electron bunches passing through dielectric-lined waveguide structures (capillaries)~\cite{OShea16}.
%Segmented dielectric-lined waveguides were recently offered to manipulate the longitudinal phase space of the bunch~\cite{Mayet2020}.
%Inversely, dielectric capillaries similar to those used for wakefield acceleration can be in turn utilized for 
and development of high-power narrow-band radiation sources including those for Terahertz (THz) frequencies~\cite{GTAB14, IGTT14, WangAntipov2018, AntipovXiang2020}.
Typical structure for mentioned application is a dielectric capillary -- a circular waveguide with a dielectric layer and axial channel for bunch passage.
Recently, a promising three-layer modification of mentioned capillary has been offered which essentially reduces the width of the main radiation lobe and therefore enhances considerably the radiated power~\cite{Jiang2020}.  

For further development of the discussed topics a rigorous approach allowing analytical investigation of both radiation from such open-ended capillary and its excitation by external source (bunch or electromagnetic pulse) would be extremely useful.
In our recent papers~\cite{GVT2021, GV2022}, we have presented an efficient rigorous method for solving circular open-ended waveguide diffraction problems and illustrated this method using the case of uniform dielectric filling and a two-layer lining of the waveguide.
Here we deal with more complicated geometry offered in~\cite{Jiang2020} and internal excitation by single waveguide mode.
The presented technique can be directly applied to the radiation of CR wakefield generated behind the moving charge in the form of a slow waveguide mode.
Moreover, presented rigorous solution can be potentially extended to a beam-driven case (similar to how it has been done for ``embedded'' structures~\cite{GTVGA19}).

Despite the aforementioned practical importance, the present paper also contributes to the development of rigorous diffraction theory since it deals with a canonical problem, i.e. relatively simple geometric structure (so called ``canonical structure'') excited by simple free or guided wave.
%Probably the most famous canonical problems are edge and wedge diffraction problems which have resulted in both outstanding mathematical approaches and the family of composite diffraction methods based on the locality principles~\cite{Nethercote2020, AssierShanin2  018}. 
%In the context of the present paper, 
A series of related problems connected with an open end discontinuity~\cite{Weinb, Mittrab, WilliamsLighthill56, VZh78, Johnson80, Kobayashi91, KobayashiSawai92, Kobayashi97, Gupta97, Kuryliak2000, Kobayashi2004, Hames2004, Hames2005, Cicchetti2008, GTV17, Buyukaksoy2007, Buyukaksoy2008, HaciveliogluBuyukaksoy2009, TayyarBuyukaksoy2011} or a cross-section discontinuity~\cite{Zaki83, Zaki88} in waveguides and resonators can be mentioned.
%Separately one should note a series of papers dealt with canonical structures formed by coaxial waveguides~\cite{Buyukaksoy2007, Buyukaksoy2008, HaciveliogluBuyukaksoy2009} and especially the paper~\cite{TayyarBuyukaksoy2011} where a method similar to that used in the present paper has been utilized.
%A class of more complicated canonical geometries with open-ended parallel-plate dielectric-loaded waveguide having a wedge-shaped flange have been also considered recently~\cite{Daniele2017_1, Daniele2017_2, Daniele2019}.
However, the diffraction problem with a canonical structure discussed in this paper has not been investigated rigorously up to now.

\section{Problem formulation and general solution}

We consider an open-ended semi-infinite cylindrical waveguide with radius $ d $ lined with a dielectric $\varepsilon>1$ of thickness $ a - b $ and having a layer of thickness $ d - a $ made of dielectric $\epsilon>1$ near the waveguide wall, see Fig.~\ref{fig:geom} (cylindrical frame $\rho, \varphi, z$ is used).
Both the region outside the waveguide ($ z > 0 $ and $ z < 0 $, $ \rho > d $) and the inner channel ($ z < 0 $, $ \rho < b $) are filled with vacuum. 
Waveguide walls have an ideal electric conductivity (PEC). 
The method used for solution is the same as in~\cite{GVT2021, GV2022}. 

A $\varphi$-symmetric TM problem is considered in the harmonic regime with time dependence in the form
\begin{equation}
H_{ \varphi }(\rho, z, t) = H_{\omega \varphi }(\rho, z) \exp( -i \omega t ).
\end{equation}
This problem is formulated for the magnitude $ H_{\omega \varphi } $, other nonzero field components can be derived as follows: 
\begin{equation}
\label{eq:ErhoEz}
E_{ \omega \rho  } = \frac{ 1 } {i k_0 \tilde{\varepsilon} } \frac{ \partial H_{\omega \varphi } }{\partial z }, 
\quad
E_{ \omega z  } = \frac{ i } { k_0 \varepsilon }
\left( \frac{ H_{\omega \varphi } }{ \rho } + \frac{ \partial H_{\omega \varphi } }{\partial \rho } \right), 
\end{equation}
where $\tilde{\varepsilon} = 1$, $\varepsilon$ or $\epsilon$ depending on the region.
In particular, we have 
$ E_{ \omega z } = 0 $
for 
$ \rho = d $, 
$ z < 0 $.

We suppose that single symmetrical $T{{M}_{0l}}$ waveguide mode is incident on the orthogonal open end:
%(see~\cite{GrigTVA18})
%%%%%%%%%%%%%%%%%%%%%%%%%%%%%%
\begin{figure*}
\centering
\includegraphics[width=0.75\textwidth]{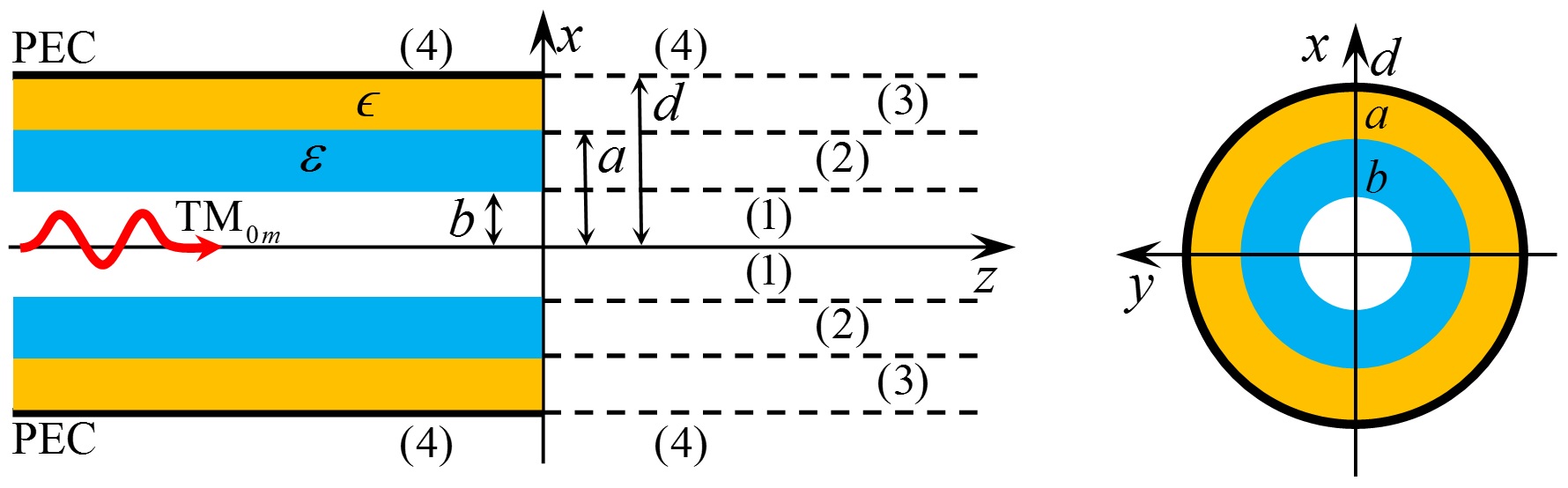}
\caption{\label{fig:geom} Geometry of the problem and main notations.}
\end{figure*}
%%%%%%%%%%%%%%%%%%%%%%%%%%%%%%
%
%
\begin{equation}
\label{eq:Hphii}
H_{\omega \varphi }^{(i)}
=
M^{(i)} e^{ i k_{ z l } z }
\left\{
\begin{aligned}
&\left. J_1 ( \rho \sigma_l ) \right/ \sigma_l \quad \text{for  } \rho < b, \\
& \frac{J_1(b\sigma_l)}{\sigma_l}\frac{\psi_l(\rho)}{\psi_l(b)}
\,\,\, \text{for  } b < \rho < a, \\
& \frac{J_1(b\sigma_l)}{\sigma_l}\frac{\psi_l(a)}{\psi_l(b)}\frac{\psi_{0l}(\rho)}{\psi_{0l}(a)}
\,\,\,\, \text{for  } a < \rho < d,
\end{aligned}
\right.
\end{equation} 
where
$ M^{(i)} $ is an arbitrary amplitude constant,
\begin{equation}
\label{eq:psi0}
\begin{aligned}
\psi_{0m}(\xi) &= J_1(\xi \chi_m) Y_0(d\chi_m) - N_1(\xi \chi_m) J_0(d\chi_m), \\
\psi_{m}(\xi) &= J_1(\xi s_m) \varphi_{2m} - N_1(\xi s_m) \varphi_{1m}, \\
\varphi_{1m} &= \sigma_m J_0(b\sigma_m)J_1(bs_m) - s_m\varepsilon^{-1}J_1(b\sigma_m)J_0(bs_m), \\
\varphi_{2m} &= \sigma_m J_0(b\sigma_m)Y_1(bs_m) - s_m\varepsilon^{-1}J_1(b\sigma_m)Y_0(bs_m),
\end{aligned}
\end{equation}
$ J_{ \nu } $ and $ Y_{ \nu } $ are Bessel and Neumann functions of $\nu$-th order, correspondingly, transverse wave numbers $ \sigma_m $, $ s_m $ and $\chi_m$ are determined by the following dispersion equation
\begin{equation}
\label{eq:disp1}
\mathrm{det}\hat{M}=0,
\end{equation}
\begin{equation}
\label{eq:disp1}
\hat{M} {=}
\left(
\begin{array}{cccc}
\varphi_{1m} & \varphi_{2m} & 0 & 0 \\
J_1(as_m) & Y_1(as_m) & -J_1(a\chi_m) & -Y_1(a\chi_m) \\
\frac{s_mJ_0(as_m)}{\varepsilon} & \frac{s_mY_0(as_m)}{\varepsilon} & \frac{-\chi_m J_0(a\chi_m)}{\epsilon} & \frac{-\chi_m Y_0(a\chi_m)}{\epsilon} \\
0 & 0 & J_0(d\chi_m) & Y_0(d\chi_m)
\end{array}
\right),
\end{equation}
longitudinal wave number $ k_z $ is connected with $ \sigma_m $, $ s_m $ and $\chi_m$ as follows:
\begin{equation}
\label{eq:lwn}
\begin{aligned}
k_{ z m } &= \sqrt{ k_0^2 - \sigma_m^2 } = \sqrt{ k_0^2 \varepsilon - s_m^2 } = \\
& = \sqrt{ k_0^2 \epsilon - \chi_m^2 },
\quad
\mathrm{Im} k_{ z m } > 0,
\end{aligned}
\end{equation}
$ k_0 = \omega / c + i \delta$
(%
$ \delta \to 0 $,
which is equivalent to infinitely small dissipation an all areas), 
$ c $ is the light speed in vacuum.
From~\eqref{eq:lwn} one can express both $ s_m $ and $ \chi_m $ through $ \sigma_m $ and obtain the dispersion relation with respect to a single variable $ \sigma_m $.
%Note that $ \sigma_0 = 0 $ is the solution of the dispersion equation~\eqref{eq:disp} if the following condition for the frequency holds: 
%$ 2 \varepsilon \psi_0( s_0 ) = b s_0 \psi_1( s_0 )$.
%Corresponding mode propagates with the speed of light $ c $ because $ k_{ z 0 } = k_0 $.
Note that Eq.~\eqref{eq:Hphii} transforms to the corresponding mode of a two-layer waveguide~\cite{GV2022} for $\epsilon \to \varepsilon$.

The reflected field in the area inside the waveguide ($ z < 0 $, $ \rho < a $) is decomposed into a series of TM modes propagating in the opposite direction:
\begin{equation}
\label{eq:r_fields}
H_{\omega \varphi }^{(r)}
=
\sum\limits_{m=0}^{\infty}
M_m e^{ -i k_{ z m } z }
\left\{
\begin{aligned}
&\left. J_1 ( \rho \sigma_m ) \right/ \sigma_m \quad \text{for  } \rho < b, \\
& \frac{J_1(b\sigma_m)}{\sigma_m}\frac{\psi_m(\rho)}{\psi_m(b)}
\,\,\, \text{for  } b < \rho < a, \\
& \frac{J_1(b\sigma_m)}{\sigma_m}\frac{\psi_m(a)}{\psi_m(b)}\frac{\psi_{0m}(\rho)}{\psi_{0m}(a)} \\
&\quad \quad \quad \quad \quad \text{for  } a < \rho < d,
\end{aligned}
\right.
\end{equation}
where $\{ M_m \}$ are unknown ``reflection coefficients'' that should be determined.
Note that since both the structure and the excitation~\eqref{eq:Hphii} are uniform in $\varphi$ then only symmetric TM modes are taken into account in the reflected field.
%~\eqref{eq:r_fields}~\cite{Zaki83, Zaki88, TayyarBuyukaksoy2011}.

The area outside the waveguide is divided into three subareas ``1'', ``2'', ``3'' and ``4'' (see Fig.~\ref{fig:geom}), where the field is described by Helmholtz equation:
\begin{equation}
\label{eq:Helmholtz}
\left[
\frac{ \partial^{ 2 } }{ \partial z^2 }+\frac{ \partial^2 }{ \partial \rho^2 }+
\frac{ 1 }{ \rho } \frac{ \partial }{\partial \rho } +
\left( k_0^2 - \frac{ 1 }{ \rho^2 } \right)
\right]
H_{\omega \varphi}^{(1,2,3,4)}=0.
\end{equation}
We introduce functions ${{\Psi }_{\pm }}(\rho ,\alpha )$ (hereafter subscripts $\pm $ mean that function is holomorphic and free of poles and zeros in areas $\operatorname{Im}\alpha >-\delta$ and $\operatorname{Im}\alpha <\delta$, correspondingly):
\begin{equation}
\label{eq:plustrans}
\Psi _{+}^{(1,2,3,4)}(\rho ,\alpha )= {{(2\pi )}^{-1}}\int_{0}^{\infty }{dzH_{\omega \varphi}^{(1,2,3,4)}(\rho ,z){{e}^{i\alpha z}}},
\end{equation}
\begin{equation}
\label{eq:minustrans}
\Psi _{-}^{(4)}(\rho ,\alpha )={{(2\pi )}^{-1}}\int_{-\infty }^{0}{dzH_{\omega \varphi}^{(4)}(\rho ,z){{e}^{i\alpha z}}},
\end{equation}
and similar transforms of
$ E_{\omega z}^{(1, 2, 3, 4)} $,
for example,
\begin{equation}
\label{eq:Phi}
\Phi_{+}^{(1,2,3,4)}(\rho ,\alpha )={{(2\pi )}^{-1}}\int_{0 }^{\infty}{dz \frac{ k_0 }{ i } E_{\omega z}^{(1,2,3,4)}(\rho ,z){{e}^{i\alpha z}}}.
\end{equation}
From~\eqref{eq:Phi} and \eqref{eq:ErhoEz} we have the following relation between $\Phi$ and $\Psi$:
\begin{equation}
\label{eq:Psi2Phi}
\Phi_+^{ ( 1, 2, 3, 4 ) }( \rho, \alpha )
=
\frac{ \Psi_+^{ ( 1, 2, 3, 4 ) }( \rho, \alpha ) }{ \rho } + \frac{ \partial \Psi_+^{ ( 1, 2, 3, 4 ) }( \rho, \alpha ) }{\partial \rho },
\end{equation}
and the same relation between $ \Phi_-^{ ( 4 ) }( \rho, \alpha ) $ and $ \Psi_-^{ ( 4 ) }( \rho, \alpha ) $. 

From~\eqref{eq:Helmholtz} we obtain
\begin{equation}
\label{eq:Psi}
\left( \frac{{{\partial }^{2}}}{\partial {{\rho }^{2}}} {+} \frac{1}{\rho }\frac{\partial }{\partial \rho } {+} {{\kappa }^{2}} {-} \frac{1}{\rho^2} \right)
\left\{ \begin{matrix}
   \Psi _{+}^{(1,2,3)}  \\
   \Psi _{-}^{(4)}{+}\Psi _{+}^{(4)}  \\
\end{matrix}
\right\}{=}
\left\{
\begin{matrix}
   {{F}^{(1,2,3)}}  \\
   0  \\
\end{matrix} \right\},
\end{equation}
\begin{equation}
2 \pi F^{ ( 1, 2, 3 ) } = 
{ \left. {\partial H_{\omega \varphi}^{(1,2,3)}}/{\partial z} \right| }_{ z=0_+ } - {\left. i\alpha H_{\omega \varphi}^{(1,2,3)} \right|}_{ z=0_+ },
\end{equation}
where 
\begin{equation}
\kappa = \sqrt{k_{0}^{2}-{{\alpha }^{2}}}, 
\quad
\operatorname{Im}\kappa >0.
\end{equation}
Equation~\eqref{eq:Psi} is obtained in a similar way as Eq.~(9) in~\cite{GVT2021}.

Functions ${{F}^{(1,2,3)}}$ are determined using continuity of
$ E_{ \omega \rho } $ and $ H_{ \omega \varphi} $ for $ z = 0 $, $\rho < d $ (see~\cite{GVT2021} for details), 
therefore
\begin{equation}
\label{eq:F1calc}
\begin{aligned}
2\pi{{F}^{(1)}} &= 
i \left[
\vphantom{\sum\nolimits_{m=1}^{\infty }} 
\left( k_{ z l } - \alpha \right)
M^{(i)} \left. J_1 ( \rho \sigma_l ) \right/ \sigma_l \right. \\
&\left.
-\sum\nolimits_{ m = 1 }^{ \infty } 
\left( k_{ z m } + \alpha \right)
M_m \left. J_1 ( \rho \sigma_m ) \right/ \sigma_m
\right], 
\end{aligned}
\end{equation}
\begin{equation}
\label{eq:F2calc}
\begin{aligned}
2 \pi F^{ ( 2 ) } 
&= 
i \left[
M^{(i)}
\left( \frac{ k_{ z l } }{ \varepsilon } - \alpha \right)
\frac{ J_1 ( b \sigma_l ) }{ \sigma_l } \frac{\psi_l(\rho)}{\psi_l(b)} \right. \\
&-\sum\limits_{m=1}^{\infty } M_m \left( \frac{ k_{zm} }{\varepsilon} + \alpha \right)
\frac{ J_1 ( b \sigma_m ) }{ \sigma_m } \frac{\psi_m(\rho)}{\psi_m(b)}
\left.
\vphantom{ \left( \frac{ k_{ z l } }{ \varepsilon } - \alpha \right) }
\right], 
\end{aligned}
\end{equation}
\begin{equation}
\label{eq:F3calc}
\begin{aligned}
2 \pi F^{ ( 3 ) } 
&= 
i \left[
M^{(i)}
\left( \frac{ k_{ z l } }{ \epsilon } - \alpha \right)
\frac{ J_1 ( b \sigma_l ) }{ \sigma_l } \frac{\psi_l(a)}{\psi_l(b)} \frac{\psi_{0l}(\rho)}{\psi_{0l}(a)} \right. \\
&-\sum\limits_{m=1}^{\infty } M_m \left( \frac{ k_{zm} }{\epsilon} + \alpha \right)
\frac{ J_1 ( b \sigma_m ) }{ \sigma_m } \frac{\psi_m(a)}{\psi_m(b)}\frac{\psi_{0m}(\rho)}{\psi_{0m}(a)}
\left.
\vphantom{ \left( \frac{ k_{ z l } }{ \varepsilon } - \alpha \right) }
\right]. 
\end{aligned}
\end{equation}

General solution of Eq.~\eqref{eq:Psi} has the form (see~\cite{GVT2021} for details):
\begin{align}
\label{eq:gen_sol_1}
\Psi_+^{ ( 1 ) }( \rho, \alpha ) 
&=
C^{(1)} J_1 ( \rho \kappa ) + \Psi_p^{ ( 1 ) }( \rho, \alpha ), \\
\label{eq:gen_sol_2}
\Psi_+^{ ( 2 ) }( \rho, \alpha ) &= C^{ ( 2 ) }_1 J_1 ( \rho \kappa ) \nonumber \\
&+ C^{ ( 2 ) }_2 Y_1( \rho \kappa ) + \Psi_p^{ ( 2 ) }( \rho, \alpha ), \\
\label{eq:gen_sol_3}
\Psi_+^{ ( 3 ) }( \rho, \alpha ) &= C^{ ( 3 ) }_1 J_1 ( \rho \kappa ) \nonumber \\
&+ C^{ ( 3 ) }_2 Y_1( \rho \kappa ) + \Psi_p^{ ( 3 ) }( \rho, \alpha ), \\
\label{eq:gen_sol_4}
\Psi_-^{ ( 4 ) }( \rho, \alpha ) &+ \Psi_+^{ ( 4 ) }( \rho, \alpha ) = C^{ ( 4 ) } H_1^{ ( 1 ) }( \rho \kappa ),
\end{align}
where 
$H^{ ( 1 ) }_{ \nu }$
is a Hankel function of the first kind of $\nu$-th order,
$ C^{ ( 1, 4 ) } $ and $ C^{ ( 2, 3 ) }_{ 1, 2 } $ are unknown coefficients.

Particular solutions of the inhomogeneous equations $ \Psi_p^{ ( 1, 2, 3 ) } $ have the form (see~\cite{GVT2021} for details)
\begin{equation}
\label{eq:partsol1}
\Psi_p^{ ( 1 ) }( \rho, \alpha )
=
\frac{ i M^{(i)} }{ 2 \pi \sigma_l }
\frac{ J_1 \left( \rho \sigma_l \right) }{ k_{zl} + \alpha } 
-
\sum\limits_{m=1}^{\infty }
\frac{ i M_m }{ 2 \pi \sigma_m }
\frac{ J_1 \left( \rho \sigma_m \right) }{ k_{zm} - \alpha },
\end{equation}
\begin{equation}
\begin{aligned}
\label{eq:partsol2}
&\Psi _{p}^{(2)}( \rho, \alpha )
=
\frac{ i M^{ ( i ) } }{ 2 \pi }
\frac{ \frac{ k_{zl} }{ \varepsilon } - \alpha }{ \kappa^2 - s_l^2 } 
\frac{ J_1 ( b \sigma_l ) }{ \sigma_l } \frac{\psi_l(\rho)}{\psi_l(b)} \\
&-
\sum\limits_{m=1}^{\infty }
\frac{ i M_m }{ 2 \pi }
\frac{  \frac{ k_{ z m } }{ \varepsilon } + \alpha }{ \kappa^2 - s_m^2 }
\frac{ J_1 ( b \sigma_m ) }{ \sigma_m } \frac{\psi_m(\rho)}{\psi_m(b)}, \\
\end{aligned}
\end{equation}
\begin{equation}
\begin{aligned}
\label{eq:partsol3}
&\Psi _{p}^{(3)}( \rho, \alpha )
=
\frac{ i M^{ ( i ) } }{ 2 \pi }
\frac{ \frac{ k_{zl} }{ \epsilon } - \alpha }{ \kappa^2 - \chi_l^2 } 
\frac{ J_1 ( b \sigma_l ) }{ \sigma_l } \frac{\psi_l(a)}{\psi_l(b)} \frac{\psi_{0l}(\rho)}{\psi_{0l}(a)} \\
&-
\sum\limits_{m=1}^{\infty }
\frac{ i M_m }{ 2 \pi }
\frac{  \frac{ k_{ z m } }{ \epsilon } + \alpha }{ \kappa^2 - \chi_m^2 }
\frac{ J_1 ( b \sigma_m ) }{ \sigma_m } \frac{\psi_m(a)}{\psi_m(b)} \frac{\psi_{0m}(\rho)}{\psi_{0m}(a)}. \\
\end{aligned}
\end{equation}
From Eq.~\eqref{eq:Psi2Phi} one obtains corresponding form of general solution for $\Phi$:
\begin{align}
\label{eq:gen_sol_Phi_1}
\Phi_+^{ ( 1 ) }( \rho, \alpha ) &= i k_0^{ -1 } C^{(1)} \kappa J_0 ( \rho \kappa ) + \Phi_p^{ ( 1 ) }( \rho, \alpha ), \\
\label{eq:gen_sol_Phi_2}
\Phi_+^{ ( 2 ) }( \rho, \alpha ) &= i k_0^{-1} C^{ ( 2 ) }_1 \kappa J_0 ( \rho \kappa ) \nonumber \\
&+ i k_0^{ -1 } C^{ ( 2 ) }_2 \kappa J_0( \rho \kappa ) + \Phi_p^{ ( 2 ) }( \rho, \alpha ), \\
\label{eq:gen_sol_Phi_3}
\Phi_+^{ ( 3 ) }( \rho, \alpha ) &= i k_0^{-1} C^{ ( 3 ) }_1 \kappa J_0 ( \rho \kappa ) \nonumber \\
&+ i k_0^{ -1 } C^{ ( 3 ) }_2 \kappa J_0( \rho \kappa ) + \Phi_p^{ ( 3 ) }( \rho, \alpha ), \\
\label{eq:gen_sol_Phi_4}
\Phi_-^{ ( 4 ) }( \rho, \alpha ) &+ \Phi_+^{ ( 4 ) }( \rho, \alpha ) = i k_0^{ -1 } C^{ ( 4 ) } \kappa H_0^{ ( 1 ) }( \rho \kappa ),
\end{align}
where
\begin{equation}
\label{eq:partsol_Phi1}
\Phi_p^{ ( 1 ) }( \rho, \alpha )
=
\frac{ - M^{(i)} }{ 2 \pi k_0 }
\frac{ J_0 ( \rho \sigma_l ) }{ k_{ z l } + \alpha } 
+
\sum\limits_{m=1}^{\infty }
\frac{ M_m }{ 2 \pi k_0 }
\frac{ J_0 \left( \rho \sigma_m \right) }{ k_{zm} - \alpha },
\end{equation}
\begin{equation}
\begin{aligned}
\label{eq:partsol_Phi2}
&\Phi_p^{ ( 2 ) }( \rho, \alpha )
=
\frac{ - M^{ ( i ) } }{ 2 \pi k_0 }
\frac{ \frac{ k_{zl} }{ \varepsilon } - \alpha }{ \kappa^2 - s_l^2 } 
\frac{ s_l J_1 ( b \sigma_l ) }{ \sigma_l } \frac{\phi_l(\rho)}{\phi_l(b)} \\
&+
\sum\limits_{m=1}^{\infty }
\frac{ M_m }{ 2 \pi k_0 }
\frac{ \frac{ k_{ z m } }{ \varepsilon } + \alpha }{ \kappa^2 - s_m^2 }
\frac{ s_m J_1 ( b \sigma_m ) }{ \sigma_m } \frac{\phi_m(\rho)}{\phi_m(b)}, \\
\end{aligned}
\end{equation}
\begin{equation}
\begin{aligned}
\label{eq:partsol_Phi3}
&\Phi_p^{ ( 3 ) }( \rho, \alpha )
=
\frac{ - M^{ ( i ) } }{ 2 \pi k_0 }
\frac{ \frac{ k_{zl} }{ \epsilon } - \alpha }{ \kappa^2 - \chi_l^2 } 
\frac{ J_1 ( b \sigma_l ) }{ \sigma_l } \frac{\psi_l(a)}{\psi_l(b)} \frac{\chi_l \phi_{0l}(\rho)}{\phi_{0l}(a)}\\
&+
\sum\limits_{m=1}^{\infty }
\frac{ M_m }{ 2 \pi k_0 }
\frac{ \frac{ k_{ z m } }{ \varepsilon } + \alpha }{ \kappa^2 - \chi_m^2 }
\frac{ J_1 ( b \sigma_m ) }{ \sigma_m } \frac{\psi_m(a)}{\psi_m(b)} \frac{\chi_m \phi_{0m}(\rho)}{\phi_{0m}(a)}, \\
\end{aligned}
\end{equation}
where
\begin{equation}
\label{eq:phi0}
\begin{aligned}
\phi_{0m}(\rho) &= J_0(\rho \chi_m) Y_0(d \chi_m) - N_0(\rho \chi_m) J_0(d\chi_m), \\
\phi_{m}(\rho) &= J_0(\rho s_m) \varphi_{2m} - N_0(\rho s_m) \varphi_{1m}, 
\end{aligned}
\end{equation}
and
$ \Phi_p^{ ( 3 ) }( d, \alpha ) = 0 $ since $\phi_{0m}(d) = 0$. 

Boundary condition $ E_{\omega z} = 0 $ for $ \rho = d $, $ z < 0 $ results in $ \Phi_-^{(4)}( d, \alpha ) = 0 $, and we obtain from~\eqref{eq:gen_sol_Phi_4}:
\begin{equation}
\label{eq:C4}
C^{ (4 ) }
= \frac{ k_0 \Phi_+^{ ( 4 ) }( d, \alpha ) }
{ i \kappa H_1^{ ( 1 ) }( d \kappa ) },  
\end{equation}
therefore from Eq.~\eqref{eq:gen_sol_4}
\begin{equation}
\label{eq:WHFore}
\Psi_+^{ ( 4 ) }( d, \alpha ) + \Psi_-^{ ( 4 ) }( d, \alpha ) 
= 
\frac{ k_0 \Phi_+^{(4)}(a,\alpha) H_1^{ ( 1 ) }( d \kappa ) }
{ i \kappa H_1^{ ( 1 ) }( d \kappa ) }.
\end{equation}

To obtain Wiener-Hopf-Fock equation one should express the term $ \Psi_+^{ ( 4 ) }( d, \alpha ) $ in Eq.~\eqref{eq:WHFore} through the $ \Phi_+^{(4)}(d,\alpha) $. 
This can be done using the continuity conditions for $ \rho = b, a, d $, $ z > 0 $ (see~\cite{GV2022} for details): 
%$ H_{ \omega \varphi }^{(3)}( a, z ) = H_{ \omega \varphi }^{(2)}( a, z ) $ 
%and 
%$ E_{ \omega z }^{(3)}( a, z ) = E_{ \omega z }^{(2)}( a, z ) $, 
%therefore  
%$ \Psi_+^{(3)}( a, \alpha ) = \Psi_+^{(2)}( a, \alpha ) $
%and
%$ \Phi_+^{(3)}( a, \alpha ) = \Phi_+^{(2)}( a, \alpha ) $.
%Excluding the constant $ C_1^{(2)}$ from Eqs.~\eqref{eq:gen_sol_2}, \eqref{eq:gen_sol_Phi_2} we have:
%%
%\begin{equation}
%\label{eq:BCa}
%\begin{aligned}
%&C_2^{ ( 2 ) } 
%=
%\left[
%\frac{ i \kappa }{ k_0 } 
%J_0( a \kappa )
%\left(
%\Psi_p^{ ( 2 ) }( a, \alpha ) - \Psi_+^{ ( 3 ) }( a, \alpha )
%\right) \right. \\
%&-
%\left.
%J_1( a \kappa )
%\left(
%\Phi_p^{ ( 2 ) }( a, \alpha ) - \Phi_+^{ ( 3 ) }( a, \alpha ) 
%\right)
%\vphantom{ \frac{ i \kappa }{ k_0 } }
%\right]
%\frac{ \pi a k_0 }{ 2 i }.
%\end{aligned}
%\end{equation}
%%
%Similarly, continuity conditions for $ \rho = b $, $ z > 0 $ give:
%%
%%
%\begin{equation}
%\label{eq:BCb}
%\begin{aligned}
%&C_2^{(2)}
%=
%\left[
%\frac{ i \kappa }{ k_0 }
%J_0( b \kappa )
%\left(
%\Psi_p^{ ( 2 ) }( b, \alpha ) - \Psi_p^{ ( 1 ) }( b, \alpha )
%\right) \right. \\
%&-
%\left.
%J_1( b \kappa )
%\left(
%\Phi_p^{ ( 2 ) }( b, \alpha ) - \Phi_p^{ ( 1 ) }( b, \alpha )
%\right)
%\vphantom{ \frac{ i \kappa }{ k_0 } }
%\right]
%\frac{ \pi b k_0 }{ 2 i }.
%\end{aligned}
%\end{equation}
%%
%Combining Eqs.~\eqref{eq:BCa} and \eqref{eq:BCb} we obtain the required relation:
%
\begin{equation}
\label{eq:PsiPlus4}
\begin{aligned}
\Psi_+^{ ( 4 ) }( d, \alpha )
&=
\frac{ k_0 \Phi_+^{ ( 4 ) }( d, \alpha ) J_1( d \kappa ) }
{ i \kappa J_0( d \kappa ) }
+
\Psi_p^{ ( 3 ) }( d, \alpha ) \\
- \frac{ k_0 }{ i \kappa d J_0(d \kappa) }
&\left[
b J_1( b\kappa) \left( \Phi_p^{(1)}(b,\alpha) - \Phi_p^{(2)}(b,\alpha) \right) \right. \\ 
&+ a J_1( a \kappa ) \left( \Phi_p^{ ( 2 ) }( a, \alpha ) - \Phi_p^{ ( 1 ) }( a, \alpha ) \right) \\
&- \frac{i b \kappa J_0( b \kappa )}{k_0} \left( \Psi_p^{ ( 1 ) }( b, \alpha ) - \Psi_p^{ ( 2 ) }( b, \alpha ) \right) \\
&- \left. \frac{i a \kappa J_0( a \kappa )}{k_0} \left( \Psi_p^{ ( 2 ) }( a, \alpha ) - \Psi_p^{ ( 3 ) }( a, \alpha ) \right) \right]. 
\end{aligned}
\end{equation}

Similar to~\cite{GVT2021, GV2022}, % 
%The following important note should be made here.
%It can be checked that the right-hand side of~\eqref{eq:PsiPlus3} is free from pole singularities for both $ \alpha = \pm k_{ z m } $ and $ \alpha = \xi_m $, $ m = 1, 2, \ldots $, where
%%
%\begin{equation}
%\label{eq:xim}
%\xi_m = \sqrt{ k_0^2 - s_m^2 },
%\quad
%\mathrm{ Im }\xi_m > 0,
%\end{equation}
%%
%($ \xi_m $ satisfies the equation $ \kappa^2 = s_m^2 $). 
%However, 
the right-hand side of~\eqref{eq:PsiPlus4} formally possesses pole singularity for 
$ \alpha = \alpha_m $:
\begin{equation}
\label{eq:alpham}
\alpha_m = \sqrt{ k_0^2 - j_{ 0 m }^2 d^{-2} },
\quad
\mathrm{Im} \alpha_m > 0,
\end{equation}
where $ j_{ 0 m } $ 
is the $m$-th zero of Bessel function $ J_0 $ ($ \alpha_m $ are longitudinal wavenumbers of vacuum waveguide of radius $ d $).
However, the function determined by Eq.~\eqref{eq:PsiPlus4} should be regular in the area $ \mathrm{Im}\alpha>-\delta $.
Therefore, this pole singularity at the right-hand side should be eliminated and we obtain the following requirement:
\begin{align}
\label{excludepoles}
& \Phi_+^{ ( 4 ) }( d,  \alpha_p )
J_1( j_{ 0 p } )
=
\frac{ 1 }{ d }
\left\{\vphantom{\left[ \Phi_p^{(1)}(b,\alpha_p) - \Phi_p^{(2)}(b,\alpha_p) \right]}
b J_1( b j_{0p}/d) \right. \\
&\times \left[ \Phi_p^{(1)}(b,\alpha_p) - \Phi_p^{(2)}(b,\alpha_p) \right] \\ 
&+ a J_1( a j_{0p}/d ) \left[ \Phi_p^{ ( 2 ) }( a, \alpha_p ) - \Phi_p^{ ( 1 ) }( a, \alpha_p ) \right] \\
&- \frac{i b j_{0p} J_0( b j_{0p}/d )}{d k_0} \left[ \Psi_p^{ ( 1 ) }( b, \alpha_p ) - \Psi_p^{ ( 2 ) }( b, \alpha_p ) \right] \\
&- \left. \frac{i a j_{0p} J_0( a j_{0p}/d )}{d k_0} \left[ \Psi_p^{ ( 2 ) }( a, \alpha_p ) - \Psi_p^{ ( 3 ) }( a, \alpha_p ) \right] \right\}.
\end{align}
%     
%where 
%%
%\begin{align}
%&\eta_m( \alpha )
%=
%\frac{ J_1 ( b \sigma_m ) }{ \kappa^2 - s_m^2 } 
%\frac{ 1 }{ \sigma_m \psi_0( s_m ) } \\
%\times
%&\left[ 
%Y_0( a s_m )
%\left(
%s_m J_0( b s_m ) \frac{ J_1( b \kappa ) }{ \kappa } - J_0( b \kappa ) J_1( b s_m ) 
%\right)
%\right. \label{eq:etaM} \nonumber \\
%&-
%Y_0( b s_m )
%\left(
%s_m J_0( a s_m ) \frac{ J_1( b \kappa ) }{ \kappa } - J_0( a \kappa ) J_1( b s_m ) 
%\right) \nonumber \\
%&+
%\left.
%Y_1( b s_m )
%\left(
%\vphantom{ \frac{ J_1( b \kappa ) }{ \kappa } }
%J_0( b \kappa ) J_0( a s_m ) - J_0( b s_m ) J_0( a \kappa ) 
%\right)
%\right], \nonumber
%\end{align}
%%
%%
%\begin{equation}
%\label{eq:zetaM}
%\zeta_m( \alpha )
%=
%J_0 ( b \sigma_m ) \frac{ J_1 ( b \kappa ) }{ \kappa } 
%- 
%J_0( b \kappa ) \frac{ J_1( b \sigma_m ) }{ \sigma_m }. 
%\end{equation}
%

Substituting Eq.~\eqref{eq:PsiPlus4} into Eq.~\eqref{eq:WHFore} and combining the terms proportional to $ \Phi_+^{ ( 4 ) }( d,  \alpha ) $ we obtain the following Wiener-Hopf-Fock equation:
\begin{equation}
\label{WHF1}
\frac{ 2 k_0 \Phi_{+}^{ ( 4 ) }( d, \alpha ) }
{ \kappa G(\alpha ) }
+\Psi_-^{ ( 4 ) } ( d, \alpha ) 
+
\frac{ 1 }{ d J_0(d\kappa) }
\frac{ i }{ 2 \pi }
\Pi( \alpha ) = 0,
\end{equation}
where
\begin{equation}
\label{eq:G}
G(\alpha )=\pi a\kappa J_0( a \kappa ) H_0^{ ( 1 ) } ( a \kappa ),
\end{equation}
\begin{equation}
\label{eq:Pi}
\begin{aligned}
&\Pi( \alpha )
=
b \frac{M^{ ( i ) }\zeta_{1l}(\alpha)}{k_{zl}+\alpha} - \sum\limits_{m=0}^{infty} b \frac{M_m\zeta_{1m}(\alpha)}{k_{zm}-\alpha} \\
&+ M^{ ( i ) }\frac{\frac{k_{zl}}{\varepsilon} - \alpha }{ \kappa^2 - s_l^2 } \frac{s_l J_1( b\sigma_l ) }{ \sigma_l \psi_l(b) }\eta_{1l}(\alpha) \\
&- \sum\limits_{m=0}^{\infty} M_m \frac{\frac{k_{zm}}{\varepsilon} + \alpha }{ \kappa^2 - s_m^2 } \frac{s_m J_1( b\sigma_m ) }{ \sigma_m \psi_m(b) }\eta_{1m}(\alpha) \\
&+ a M^{ ( i ) } \frac{\frac{k_{zl}}{\epsilon} - \alpha }{ \kappa^2 - \chi_l^2 } \frac{ J_1( b\sigma_l ) }{ \sigma_l } \frac{ \psi_l(a)}{ \psi_l(b) \psi_{0l}(a)}\eta_{2l}(\alpha) \\
&- \sum\limits_{m=0}^{\infty} a M_m^{ ( i ) } \frac{\frac{k_{zm}}{\epsilon} + \alpha }{ \kappa^2 - \chi_m^2 } \frac{ J_1( b\sigma_m ) }{ \sigma_m } \frac{ \psi_m(a)}{ \psi_m(b) \psi_{0m}(a)}\eta_{2m}(\alpha),
\end{aligned}
\end{equation}
\begin{equation}
\begin{aligned}
&\eta_{1m}(\alpha) = 
b\phi_{2m}\mu_{bm}(\alpha) - a\phi_{2m}\mu_{am}(\alpha) \\
&- b\phi_{1m}\nu_{bm}(\alpha) + a\phi_{1m}\nu_{am}(\alpha), \\
&\mu_{bm}(\alpha) = 
J_0(b\kappa)J_1(bs_m)/s_m - J_0(bs_m)J_1(b\kappa)/\kappa, \\
&\nu_{bm}(\alpha) =
J_0(b\kappa)Y_1(bs_m)/s_m - Y_0(bs_m)J_1(b\kappa)/\kappa, \\
&\mu_{am}(\alpha) = 
J_0(a\kappa)J_1(as_m)/s_m - J_0(as_m)J_1(a\kappa)/\kappa, \\
&\nu_{am}(\alpha) =
J_0(a\kappa)Y_1(as_m)/s_m - Y_0(as_m)J_1(a\kappa)/\kappa, \\
&\eta_{2m}(\alpha) = 
\chi_m Y_0(d\chi_m) \zeta_{2m} \\
&+ J_0(d\chi_m) \left(  \chi_m J_1(a\kappa)Y_0(a\chi_m)/\kappa - J_0(a\kappa)Y_1(a\chi_m) \right) \\
&- 2 J_0(d\kappa)(\pi a \chi_m)^{-1}, \\
&\zeta_{1m}(\alpha) =
J_0(b\sigma_m)J_1(b\kappa)/\kappa - J_0(b\kappa)J_1(b\sigma_m)/\sigma_m.
\end{aligned}
\end{equation}
%
%Performing factorization
%$ \kappa = \kappa_+ \kappa_- $
%($ \kappa_{\pm} = \sqrt{ k_0 \pm \alpha } $),
%$ G(\alpha ) = G_+(\alpha ) G_-(\alpha ) $,
%multiplying the Eq.~\eqref{WHF1} by $\kappa_- G_-$ and decomposing the function
%%
%\begin{align}
%\label{eq:S}
%S( \alpha ) = 
%\frac{ b }{ a }
%\frac{ i }{ 2 \pi }
%\Pi( \alpha )
%\frac{ \kappa_-( \alpha ) G_-( \alpha ) }{ J_0( a \kappa ) } =
%S_+( \alpha ) + S_-( \alpha ), \\
%\label{eq:Splus}
%S_+( \alpha ) = 
%\frac{ -i b }{ 2 \pi a }
%\sum\limits_{ q = 1 }^{ \infty }
%\Pi( -\alpha_q )
%\frac{ \kappa_+( \alpha_q ) G_+( \alpha_q ) j_{ 0 q } }{ a^2 \alpha_q J_1( j_{ 0 q } ) ( \alpha + \alpha_q ) },
%\end{align}
%%
%the following equation can be obtained: 
%%
%\begin{equation}
%\label{WHF2}
%\begin{aligned}
%\frac{ 2 k_0 \Phi _{+}^{ ( 3 ) } ( a, \alpha ) }{ \kappa_+( \alpha ) G_+(\alpha ) }
%&+
%S_+( \alpha ) = - \kappa_-( \alpha ) G_-(\alpha ) \Psi _-^{ ( 3 ) } ( a, \alpha ) \\
%&- S_-( \alpha ),
%\quad
%\quad
%-\delta<\mathrm{Im}\alpha < \delta.
%\end{aligned}
%\end{equation}
%%

Equation~\eqref{WHF1} is solved in a common way (see~\cite{Mittrab, GVT2021, GV2022} for details).
Omitting standard factorizations and estimations, we obtain: 
\begin{equation}
\label{WHFsol}
\begin{aligned}
&\Phi_+^{ ( 4 ) }( d, \alpha ) 
= 
\frac{ 1 }{ 4 \pi i k_0 d } \kappa_+ ( \alpha ) G_+( \alpha ) \\
&\times
\sum\limits_{ q = 1 }^{ \infty } 
\Pi( -\alpha_q )
\frac{ \kappa_+( \alpha_q ) G_+( \alpha_q ) j_{ 0 q } }{ d^2 \alpha_q J_1( j_{ 0 q } ) ( \alpha + \alpha_q ) }.
\end{aligned}
\end{equation}

To find $\{ M_m \}$, one should substitute~\eqref{WHFsol} into~\eqref{excludepoles}.
After transformations we obtain an infinite linear system of the form
\begin{equation}
\label{sys}
\sum\nolimits_{ m = 1 }^{ \infty } W_{ p m } M_m = M^{ ( i ) } w_p,
\quad
p = 1, 2, \ldots,
\end{equation}
%
%where
%%
%\begin{align}
%& W_{ p m } =
%\left( \frac{ k_{ z m } }{ \varepsilon } + \alpha_p \right)
%\eta_m( \alpha_p ) -
%\frac{ \zeta_m( \alpha_p ) }{ k_{ z m } - \alpha_p } \nonumber \\
%&+
%\frac{ J_1( j_{ 0 p } ) }{ 2 i j_{ 0 p } / a }
%\kappa_+ ( \alpha_p ) G_+( \alpha_p ) \nonumber \\
%&\times
%\sum\limits_{ q = 1 }^{ \infty }
%\left[
%\left( \frac{ k_{ z m } }{ \varepsilon } - \alpha_q \right)
%\eta_m( \alpha_q ) -
%\frac{ \zeta_m( \alpha_q ) }{ k_{ z m } + \alpha_q }
%\right] \nonumber \\
%& \times
%\frac{ \kappa_+( \alpha_q ) G_+( \alpha_q ) j_{ 0 q } }{ a^2 \alpha_q J_1( j_{ 0 q } ) ( \alpha_p + \alpha_q ) },
%\label{W}
%\end{align}
%%
%%
%\begin{align}
%& w_p =
%\left( \frac{ k_{ z l } }{ \varepsilon } - \alpha_p \right)
%\eta_l( \alpha_p ) -
%\frac{ \zeta_l( \alpha_p ) }{ k_{ z l } + \alpha_p } \nonumber \\
%&+
%\frac{ J_1( j_{ 0 p } ) }{ 2 i j_{ 0 p } / a }
%\kappa_+ ( \alpha_p ) G_+( \alpha_p ) \nonumber \\
%&\times
%\sum\limits_{ q = 1 }^{ \infty }
%\left[
%\left( \frac{ k_{ z l } }{ \varepsilon } + \alpha_q \right)
%\eta_l( \alpha_q ) -
%\frac{ \zeta_l( \alpha_q ) }{ k_{ z l } - \alpha_q }
%\right] \nonumber \\
%& \times
%\frac{ \kappa_+( \alpha_q ) G_+( \alpha_q ) j_{ 0 q } }{ a^2 \alpha_q J_1( j_{ 0 q } ) ( \alpha_p + \alpha_q ) }.
%\label{w}
%\end{align}
%% 
%For finite $ p $ and $ m \to +\infty $ we have $W_{pm}M_m \sim m^{-3/2-\tau} $ 
which is convergent and can be solved numerically using the reducing technique (see, for example,~\cite{GVT2021, GV2022} for details).

\section{EM field derivation}

When the set of coefficients $\{M_m\}$ is determined, the electromagnetic field can be calculated via the inverse transform over $\alpha$, in accordance with Eqs.~\eqref{eq:plustrans} and \eqref{eq:minustrans}:
\begin{align}
\label{eq:Hphi123}
H_{\omega\varphi}^{(1,2,3)}(\rho, z > 0) &= \int\nolimits_{-\infty}^{+\infty} \Psi^{(1,2,3)}_+(\rho,\alpha) e^{-i\alpha z} \,d\alpha, \\
\label{eq:Hphi4integral}
H_{\omega\varphi}^{(4)}(\rho, z) &= \int\nolimits_{-\infty}^{+\infty} \Psi^{(4)}(\rho,\alpha) e^{-i\alpha z} \,d\alpha.
\end{align}
performing calculations similar to those done in~\cite{GV2022}, one obtains that the field in the region $z>0$ is described by the unified formula:
\begin{equation}
\label{eq:Hzg0}
H_{\omega\varphi}^{(1,2,3)}(\rho, z>0)
= 
\sum\limits_{ q = 1 }^{ \infty }
\Pi( -\alpha_q )
\frac{ \kappa_+( \alpha_q ) G_+( \alpha_q ) j_{ 0 q } }{ d^2 \alpha_q J_1( j_{ 0 q } ) }
\frac{ L_q^+(\rho, z) }{ 2 },
\end{equation}
where $L_q^+$ is given by Eq.~(47) in~\cite{GVT2021}.

For the domain ``4'' one can also obtain the representation convenient for investigation of the far-field:
\begin{equation}
\label{eq:Hphi4}
H_{\omega\varphi}^{(4)}(\rho, z) 
= 
\frac{ 1 }{ d }
\sum\limits_{ q = 1 }^{ \infty } 
\Pi( -\alpha_{ q } )
\frac{ \kappa_+( \alpha_{ q } ) G_+( \alpha_{ q } ) j_{ 0 q } }
{ d^2 \alpha_{ q } J_1( j_{ 0 q } ) }
\frac{ I_{ q }^{ ( 4 ) } }{ 4 \pi },
\end{equation}
where integral
\begin{equation}
\label{eq:Im4}
I_{ q }^{(4)}( \rho, z )
=
\int\limits_{-\infty}^{+\infty} \!\!
\frac{ \kappa_+( \alpha ) G_+( \alpha ) H_{1}^{(1)}( \rho \kappa ) }
{ \kappa( \alpha ) H_0^{(1)}( a \kappa ) ( \alpha_{ q } + \alpha ) }
e^{ -i \alpha z }
\,d\alpha
\end{equation}
has been investigated previously (see Eq.~(41) in~\cite{GVT2021}).
%In particular, for $z>0$ we obtain the same representation for the EM field~\eqref{eq:Hzg0}.
%Finally, Eq.~\eqref{eq:H(2)fin} describes the EM field in the whole half-space $ z > 0 $.
In particular, integral~\eqref{eq:Im4} can be easily calculated asymptotically (see Eq.~(51) in~\cite{GVT2021}) and substituted into~\eqref{eq:Hphi4} to obtain far-field in the region ``4''.

\section{Conclusion\label{sec:concl}}

We have presented convenient rigorous analytical approach for calculation of various diffraction processes at the open end (with orthogonal cut) of a circular waveguide with three-layer dielectric lining.
%The obtained results have been tested by comparison with results of $S$-parameters simulations in commercial code COMSOL and an excellent agreement has been observed.
This approach can be effectively used for investigation of radiation of Cherenkov wakefield from the open-ended capillary (see Ref.~\cite{Jiang2020}) which is a promising structure for realization of beam-driven THz source with small divergence and high efficiency.
Moreover, this approach can be extended to more complicated problems.
For example, excitation by a charged particle bunch (in full formulation including both wakefield radiation and transition radiation) can be incorporated into the solution.

\section{Acknowledgements}

Author is grateful to A.M.~Altmark for fruitful discussions.
This work is supported by the Russian Science Foundation (grant No. 18-72-10137).

%%%%%%%%%%%%%%%%%%%%%%%%%%%%%%%%%%%%%%%%%%%%%%%%%%%%%%%%%%%%%%%%%%%%%%%%%%%%%%%%%%%%%%%%%%%%%

%\bibliography{SNG_Bibliography_May2022}
%

\end{document}